\documentclass[12pt]{article}
\usepackage{graphics}
\begin{document}
\begin{center}
{The Non-perturbative term of the Pseudovector Interaction \\ in the Pion-Nucleon System and the Pion Form Factor}
\end{center}
\begin{center}
{Susumu Kinpara}
\end{center}
\begin{center}
{\it Institute for Quantum Medical Science (QST)\\ Chiba 263-8555, Japan}
\end{center}
\begin{abstract}
The correction of the higher-order process for the photon-pion-pion vertex is calculated by the field theoretical method of the pion-nucleon system. 
The non-perturbative term of the pseudovector coupling interaction is included to examine the effect of the self-energy.
\end{abstract}
\section*{\normalsize{1 \quad Introduction}}
\hspace*{4.mm}
The interaction of nucleon with other particles is one of the interesting subjects and a lot of experiments makes us study the phenomena
of the related processes.
The meson-exchange model plays a decisive role in the framework of the method of the quantum fields and elucidates the validity of the various mesons
as the elements to investigate the dynamical properties.
While the calculation of the perturbative expansion is practical the result of the higher-order correction often encounters 
the difficulties of the divergence and the subtraction is necessary to compare with the experimental value.  
\\\hspace{4.mm}
Among mesons which mediate the nuclear force between nucleons the lightest one that is pion is able to reach the longest range 
and is supposed to occupy the areas around the nucleon mainly.
The pseudovector coupling interaction is the starting point to proceed the calculation of the pion-nucleon system.
Because of the derivative coupling the counter term for the mass and the field is not sufficient 
to deal with the divergences in the self-energy of the nucleon propagator.
The residual part of the divergence is removed by the additional terms which contain the self-energy 
and the fraction of two terms cancels out to obtain the finite quantity.
\\\hspace{4.mm}
The non-perturbative term is generated from the derivative on time to construct the $T$-product for the pion-nucleon-nucleon vertex part.
The improved form of the vertex is connected with the pseudoscalar coupling interaction which is common to calculate the nuclear system.
Since the difference between two interactions mentioned above is expressed by the self-energy the function form is essential to investigate the effect.
To determine it we have applied the method of the matrix inversion 
under the set of the phase-shift parameters for the $\pi$-nucleon elastic scattering $\cite{Kinpara}$. 
In the present study the model is examined whether it is consistent with the form factor of pion.
\\\hspace{4.mm}
\section*{\normalsize{2 \quad The calculation of the pion form factor}}
\hspace{4.mm}
The photon-pion system is treated by the lagrangian which consists of the electromagnetic field $A_\mu(x)$ 
and the isovector real scalar fields $\vec{\varphi}(x)$. 
The interaction is generated by the the minimal coupling as $\partial_\mu \rightarrow \partial_\mu -i e A_\mu(x)$ similar to the photon-nucleon system.
Then the general relations are useful following the method of the quantum electrodynamics for the photon-nucleon case.
It is a remarkable point that the $\sim e^2$ part of the current which is coupled with the field $A_\mu(x)$ in the equation of the motion 
is independent of the general relation. 
Therefore the method to treat the photon-nucleon-nucleon vertex is applicable to the photon-pion-pion case as seen below.
\\\hspace{4.mm}
When the scattering takes place under the electromagnetic force the higher-order of the ladder type is neglected in the beginning
and the detail of the interaction is represented by the vertex function.
For the electron-pion elastic scattering the vertex function of the photon-pion-pion part is indispensable to examine 
the effect of the extended structure of pion.
The non-perturbative relation is given as follows 
\begin{eqnarray}
(p^\prime-p) \cdot \Gamma(p^\prime,p) = \Delta(p^\prime)^{-1} - \Delta(p)^{-1}
\end{eqnarray}
which relates the vertex function $\Gamma(p^\prime,p)$ to the pion propagator $\Delta(p)$ as a function of the initial ($p$) and the final ($p^\prime$) momenta 
analogous to the Ward-Takahashi identity for the photon-electron or nucleon systems.
\\\hspace{4.mm}
The exact pion propagator $\Delta(p)$ is expressed by using the polarization function $\Pi(p)$ as
\begin{eqnarray}
\Delta(p) = \frac{1}{p^2 - m^2 - \Pi(p)}
\end{eqnarray}
\begin{eqnarray}
\Pi(p) = \sum_{n=2}^\infty \Pi_n (p^2-m^2)^n
\end{eqnarray}
under the mass and the wave function renormalization.
The expansion of $\Pi(p)$ in the series of $p^2-m^2$ is possible around the on-shell point $p^2 = m^2$ to determine the coefficients $\Pi_n$.
Using Eqs. (2) and (3) to the right-hand side in Eq. (1) the form of $\Gamma(p^\prime,p)$ is 
\begin{eqnarray}
\Gamma(p^\prime,p) = (p+p^\prime) (\, 1 - \sum_{n=2}^\infty \Pi_n \, f_n(a,b) \,) + \Gamma_0(p^\prime,p)
\end{eqnarray}
\begin{eqnarray}
f_n (a,b) = a^{n-1} \sum_{k=0}^{n-1} \sum_{l=0}^k {k\choose l} \left( \frac{b-a}{a}  \right)^l 
\end{eqnarray}
in terms of $a\equiv {p^\prime}^2-m^2$ and $b\equiv p^2-m^2$.
$\Gamma_0(p^\prime,p)$ is the additional term which suffices the relation $(p^\prime-p) \cdot \Gamma_0(p^\prime,p) = 0$.
In the actual use of $\Gamma(p^\prime,p)$ for the electron-pion elastic scattering the on-shell conditions ($a = b = 0$) give $f_n (0,0) = 0$
and only the lowest-order survives except for $\Gamma_0(p^\prime,p)$ which supplies the form factor of pion.
\\\hspace{4.mm}
To construct $\Gamma_0(p^\prime,p)$ the nucleon propagator is required and the closed loop is calculated perturbatively by using the rule of the diagram.
The anomalous interaction for the fermion loop is suitable
since the constraint for $\Gamma_0(p^\prime,p)$ is apparently fulfilled as $q^\mu \sigma_{\mu\nu} q^\nu = 0$ where $q \equiv p^\prime-p$.
By the anomalous interaction the nucleon propagator in the loop diagram is divided into two parts and the trace consists of three nucleon propagators.
Besides the difference of the sign these processes characterized by the direction of the rotation is found to give the same result.
Then it is enough to calculate only the process of the scattering by the proton in the loop diagram.  
\\\hspace{4.mm}
Following the rule of the diagram and the definition of the vertex part $I_\mu (p^\prime , p)$ in Ref. \cite{BD}
the expression in terms of the anomalous interaction by the substitution $\gamma_\mu \rightarrow \kappa \, i \, \sigma_{\mu\nu} q^\nu (2 M)^{-1}$ 
to the photon vertex is given as
\\\hspace{4.mm}
\begin{eqnarray}
I_\mu (p^\prime , p) = e \, G^2 \, \kappa \, (2 M)^{-1} q^\nu (\, I_{\mu\nu}(p^\prime,p)-I_{\nu\mu}(p^\prime,p) \,)
\end{eqnarray}
\begin{eqnarray}
I_{\mu\nu}(p^\prime,p) \equiv \int \frac{d^4 k}{i (2 \pi)^4} \frac{T_{\mu\nu}(k,p^\prime,p)}{(k^2-M^2)((p^\prime+k)^2-M^2)((p+k)^2-M^2)}
\end{eqnarray}
\begin{eqnarray}
T_{\mu\nu}(k,p^\prime,p) \equiv {\rm Tr} [\gamma_5 (\gamma\cdot k+M) \gamma_5 (\gamma\cdot (p^\prime+k)+M) \gamma_\mu \gamma_\nu (\gamma\cdot (p+k)+M)]
\end{eqnarray}
\\
in which $G \equiv 2 M f /m$ with the coupling constant of the pseudovector interaction $f \,(\approx 1)$, the nucleon and the pion masses $M$ and $m$. 
\\\hspace{4.mm}
The anti-symmetric part of $T_{\mu\nu}(k,p^\prime,p)$ results in 
\begin{eqnarray}
T_{\mu\nu}(k,p^\prime,p)-T_{\nu\mu}(k,p^\prime,p) = 4 M \, (-P_\mu q_\nu +q_\mu P_\nu)   \;\; \;\;(P \equiv p+p^\prime)
\end{eqnarray}
which is independent of $k$ and it makes the calculation of the scattering by neutron tractable.
The process is obtained by the interchange of the momenta of pion as $p^\prime \leftrightarrow -p$ and then after all by $\kappa_p \rightarrow -\kappa_n$
where $\kappa_p$ and $\kappa_n$ are the strength of the anomalous interactions for proton and neutron respectively.
Taking them into account the correction of the vertex is
\begin{eqnarray}
I_\mu(p^\prime,p) = -2 e (\kappa_p-\kappa_n)\,G^2 (P_\mu q^2 -q_\mu P\cdot q) \, I(p^\prime,p)
\end{eqnarray}
up to the $O((q^2)^0)$ term of the $k$-integral.
The $k$-integral 
\begin{eqnarray}
I(p^\prime,p) \equiv \int \frac{d^4 k}{i (2 \pi)^4} \frac{1}{(k^2-M^2)((p^\prime+k)^2-M^2)((p+k)^2-M^2)} \nonumber
\end{eqnarray}
\begin{eqnarray}
= -\frac{1}{2 (4 \pi)^2 M^2}( 1 + \frac{q^2}{12 M^2} + O((q^2)^2))
\end{eqnarray}
is used by neglecting the pion mass term ($\sim m^2$). 
\\\hspace{4.mm}
Under the on-shell condition $p^2 = {p^\prime}^2 = m^2$ of the external pions the $I_\mu(p^\prime,p)$ in Eq. (10) gives the form factor $F_\pi(q^2)$ as
\begin{eqnarray}
F_\pi(q^2) = 1 + \lambda \frac{(\kappa_p-\kappa_n)\,G^2}{(4 \pi)^2} \cdot \frac{q^2}{M^2} + O((q^2)^2)    \qquad (\lambda = 1)
\end{eqnarray}
with the factor $\lambda$ to represent the correction of the self-energy.
The charge radius of pion is $\sqrt{\langle r^2 \rangle} = \lim_{q^2 \rightarrow 0} \sqrt{6 \, d F_\pi(q^2)/d q^2} = $ 1.06$\,$fm which is
larger than the experimental value 0.672$\pm$0.008$\,$fm \cite{PDG}.
It necessitates the additional effects of the interaction to construct the $q^2$ dependence correctly.
\\\hspace{4.mm}
The photon scattering by the external pion does not contribute to the form factor because of the on-shell condition.
The conservation of the current settles the form of $I_\mu(p^\prime,p)$
while the higher-order corrections of the pion propagator to the loop diagram change the $q^2$ dependence.
Consequently the value of the charge radius shifts from the result of the lowest-order calculation through $\lambda$.
\section*{\normalsize{3 \quad The effect of the non-perturbative term }}
\hspace{4.mm}
When the self-energy of nucleon is set equal to zero the pion-nucleon-nucleon vertex is same as the pseudoscalar interaction.
Starting from the pseudovector interaction inclusion of the non-perturbative term gives the additional term expressed by the self-energy $\Sigma(p)$
\begin{eqnarray}
\gamma_5 \,\rightarrow\, \gamma_5 \, + \, \gamma_5 \, \sigma(a,b)
\end{eqnarray}
\begin{eqnarray}
\sigma(a,b) \equiv \frac{1}{2 M}(\gamma_5 \Sigma(a) \gamma_5+\Sigma(b))
\end{eqnarray}
in which the arguments $a$ and $b$ are the momenta of the final and initial nucleons respectively. 
An advantage of the procedure for dividing the interaction into two parts is that the first term in Eq. (13) is the pseudoscalar type and the result of the 
calculation is applicable to examine the correction of the second term.
\\\hspace{4.mm}
Regarding to the self-energy it is expressed as 
\begin{eqnarray}
\Sigma(p) = M c_1(p) - \gamma \cdot p \, c_2(p)
\end{eqnarray}
in terms of $c_i(p) \,(i=1,2)$ as a function of $p$.
These quantities are expanded by a series in powers of $p^2-M^2$ around the on-shell point $p^2 = M^2$
\begin{eqnarray}
c_i (p) = \sum_{n=0}^{\infty} \frac{1}{n!} \, c_i^{(n)} (p^2-M^2)^n
\end{eqnarray}
under the conditions $c \equiv c_1^{(0)} = c_2^{(0)}$ and $c_1^{(1)}-c_2^{(1)} = c \,(2M^2)^{-1}$ necessary to construct the renormalized propagator.
\\\hspace{4.mm}
Making the assumption that the momenta of the propagators in the loop integral take the values near the on-shell point 
the terms of the higher-order are neglected so that $c_i^{(n)} \approx 0 \, (n \ge 1)$.
The approximated form 
\begin{eqnarray}
\sigma(a,b) \approx \sigma_0(a,b) \equiv c \,(\, 1 + \frac{1}{2M} \, \gamma \cdot(a-b) \,)
\end{eqnarray}
is used in the present calculation.
In addition to the pion-nucleon-nucleon vertex the propagators in the loop integral possess the self-energy 
and the correction is required to understand the effect on the form factor truly. 
At present the free propagators are substituted for the exact nucleon propagators.
\\\hspace{4.mm}
Taking into account the non-perturbative term of the pion-nucleon-nucleon vertex the anti-symmetric part of $T_{\mu\nu}(k,p^\prime,p)$
is modified so as to include the effect of the self-energy
\begin{eqnarray}
T_{\mu\nu}(k,p^\prime,p)-T_{\nu\mu}(k,p^\prime,p) = 4 M \, ( -P_\mu q_\nu +q_\mu P_\nu ) \, a_1 \nonumber
\end{eqnarray}
\begin{eqnarray}
+\,a_2 \, (P_\mu k_\nu - k_\mu P_\nu) + \, a_3 \, (k_\mu q_\nu - q_\mu k_\nu)
\end{eqnarray}
\begin{eqnarray}
a_1 = 1 - c \,(\,1+\frac{c}{4} \,) \,\frac{k^2-M^2}{M^2} 
\end{eqnarray}
\begin{eqnarray}
a_2 = - \frac{4 c}{M} \, q \cdot k
\end{eqnarray}
\begin{eqnarray}
a_3 = \frac{4 c}{M} \,(\, (1+c) (k^2-M^2) - P \cdot k - m^2 \,)
\end{eqnarray}
When there is not the self-energy ($c = 0$) it is reduced to the result of the pseudoscalar coupling.
In the case of $c \neq 0$ we need the additional terms to examine the suitable sign and the magnitude of $c$ within the present approximation.
The pion mass is set equal to zero ($m^2 = 0$) for these terms ($\sim m^2/M^2 \sim {\rm 0.02}$) are much smaller than 1.
\\\hspace{4.mm}
The numerator of the $k$-integral consists of four kinds of the terms that is the constant term, $[k_\mu k_\nu] \equiv k_\mu k_\nu -g_{\mu\nu} k^2 /4$,
$\sim (k^2-M^2)$ and $\sim k_\mu (k^2-M^2)$ on the variable $k$.
The constant term has the order of the integrand $k \sim -6$ and it is convergent as has been verified for the calculation of $I(p,p^\prime)$ in Eq. (11).
Although the order of the second term is $k \sim -4$ related to the logarithmic divergence ($\sim {\rm log}\, k$) it is convergent 
because of the cancellation between two terms in $[k_\mu k_\nu]$ 
after the shift of the variable to remove the linear term in $k$ which lowers the order to $k \sim -6$.
In spite that the order of the last term is $k \sim -3$ the constant term arising from the shift of the variable as $k \rightarrow k - q z$
is also dropped by the anti-symmetric form of $k_\mu q_\nu - q_\mu k_\nu$ and does not contribute to the result. 
\\\hspace{4.mm}
Among four terms we particularly need to pay attention to the third one ($\sim (k^2-M^2)$) since it is $k \sim -4$ and results in the divergence.
In place of the method of the dimensional reguralization we choose to use the cut-off parameter $\it\Lambda$ to make the $k$-integral finite.
Then the part of the integral $J(q^2)$ is expressed by the following relation after the shift of the variable as $k \rightarrow k-p$
\begin{eqnarray}
J(q^2) \equiv \int \frac{d^4 k}{i (2 \pi)^4} \frac{1}{((p^\prime+k)^2-M^2)((p+k)^2-M^2)} \nonumber
\end{eqnarray}
\begin{eqnarray}
= \frac{1}{(4 \pi)^2} \,{\rm log} \frac{{\it\Lambda}^2}{{\it M}^2} - \frac{1}{(4 \pi)^2} \int_0^1 d z \,{\rm log}[\,1-\frac{q^2}{{\it M}^2} z (1-z)\,]
\end{eqnarray}
It depends on the cut-off $\it\Lambda$ which could include the effect of the exact form of the propagator replaced with the free propagator
to perform the $k$-integral of the loop diagram.
\\\hspace{4.mm}
Including all terms 
the factor $\lambda$ in Eq. (12) yields 
\begin{eqnarray}
\lambda = 1 + \frac{c}{2} + c \,(\,1 + \frac{c}{2}\,) \, \rm{log} \frac{{\it\Lambda}^2}{{\it M}^2} 
\end{eqnarray}
using two parameters $c$ and $\it\Lambda$.
The numerical value of $c$ is not definite and required to determine it taking account of the calculations for the other phenomena.
\\\hspace{4.mm}
Concerning to the calculations of us the expected values are divided into two regions that is roughly $c \sim 2$ and $c \sim -0.4$.
The former is associated with the electromagnetic properties of nucleon such as the magnetic moment and the electric polarizability,
the $\beta$ decay of neutron and the pion-nucleon scattering in the intermediate region.
On the other hand the latter is proper to the description of the low-energy region, for example, the low-energy parameters of the pion-nucleon elastic scattering
and the photoproduction of pion below the resonance energy.
The numerical value of $c$ is required to take the negative value so that the form factor approaches to the experimental one appropriately.
\\\hspace{4.mm}
The sign of the last term in Eq. (23) depends on the magnitude of $\it\Lambda$.
Although the contribution using $\it\Lambda \sim M$ is supposed to be small it is ascribed to the modification of the nucleon propagators.
For example the position of the energy of the $\it\Delta$-resonance acts as a typical value of $\it\Lambda$.
It is the boundary of the process with the latter value $c \sim -0.4$ as has been observed in the calculation of the photoproduction of pion
for the low-energy regions.
\\\hspace{4.mm}
The coefficients of the anomalous interactions $\kappa_p$ and $\kappa_n$ are taken from the experimental values of the magnetic moments at present.  
These parameters are also influenced by the higher-order corrections of the pion propagators 
and the self-energy in the identity of the photon-nucleon-nucleon vertex. 
\\\hspace{4.mm}
Turning off the divergent term the result of the charge radius of pion with $c \,$=$-$0.39 decreases about 10 $\%$ also larger than the experimental value 
and it is necessary to determine the cut-off parameter carefully.
To obtain the results free from the cut-off we need some improvements to the model of the calculation which would attain to understand the form factor.
For example the loop diagram is constructed by the free propagators and the non-perturbative term is approximated by the lowest-order calculation herein.
The self-energy of the non-perturbative term could be reduced effectively.
Therefore it is possible that the anomalous interactions are also changed from the values of the anomalous magnetic moments.  
\section*{\normalsize{4 \quad Summary and remarks }}
\hspace*{4.mm}
In order to examine the properties of the pion form factor the divergence raises the issue
since the counter term prescription is not applicable to the photon-pion-pion vertex.
It is fortunate that the value of the cut-off is roughly same as the nucleon mass and is not thought to give the large effect.
The convergent result could be obtained by the use of the exact propagator of nucleon which enables to make the cut-off approach to infinity.
The difficulty is also seen in the calculation of the photon-nucleon-nucleon vertex where the free nucleon propagators are used. 
Although the respective ways to remove divergences from these form factors are different 
the joint use of the finite quantities would explain the related processes such as the elastic scattering of electron by nucleon well. 

\end{document}